\documentclass[12pt]{article}
\usepackage{epsfig}
\begin{document}

\begin {center}
{\bf  {A NEW PERSPECTIVE ON MOND}}
\vskip 5mm
{\rm D.V. BUGG \footnote {email: david.bugg@stfc.ac.uk}} \\
{ Department of Physics, Queen Mary, University of London, London E1\,4NS,
UK} \\
\vskip -1mm
\end {center}

\begin{abstract}
\noindent
A novel interpretation of MOND is presented. 
For galactic data, in addition to Newtonian acceleration, there is an 
attractive acceleration peaking at Milgrom's parameter $a_0$.
The peak lies within experimental error where $a_0 = cH_0/2\pi$; 
$H_0$ is the present-time value of the Hubble constant and $c$ the 
velocity of light.
The physical interpretation of this relation and its connection to Dark 
Energy are discussed.  

\end{abstract}

\section {Introduction}
It has long been known that rotation curves of galaxies disagree with 
Newton's law.
Famaey and McGaugh have recently provided a review of all aspects
of the data, with an exhaustive list of references \cite {Famaey}.
The most relevant astrophysical data will be discussed here.

In 1983, Milgrom proposed a modification of Newtonian mechanics called MOND
(Modified Newtonian Dynamics) \cite {MilgromA}, \cite {MilgromB}.
This has three distinct features.
Firstly, the observed total acceleration $a$ depends on the Newtonian 
acceleration $g_N$ as
\begin {equation}
a = g_N/\mu (\chi); 
\end {equation}
$\mu$ is an empirical smooth function of $\chi = a/a_0$; $a_0$ is a
universal constant $\sim 1.2 \times 10^{-10}$ m s$^{-2}$ for all galaxies.
The second feature is that for vanishingly small $g_N$, 
\begin {equation}
a \to \sqrt {a_0g_N}.
\end {equation}
Thirdly, galactic dynamics are invariant under scaling of time and space for
very low $g_N$: 
\begin {equation}
(t, {\bf r}) \to \lambda (t, {\bf r}),
\end {equation}
where $\lambda$ is a scaling factor \cite {Milgroms}. 
A star with rotational velocity $v$ in equilibrium with centrifugal 
force satisfies
\begin {equation}
v^2/r = \sqrt{ \frac {a_0GM}{r^2} }; 
\end {equation}
$G$ is the gravitational constant and $M$ the galactic mass within radius $r$; 
so $r$ cancels and
\begin {equation}
v^4 = a_0 GM.
\end {equation}
This agrees well with the empirical Tully-Fisher relation 
between observed velocities at the edges of galaxies and their luminosities 
\cite {Tully}.
McGaugh showed that this relation applies over $> 5$ decades of galactic
masses from $10^6$ to $\sim 10^{12} M_\odot$ after including the mass of
gas and dust in each galaxy \cite {McGaughA}.
It is important that in MOND there is only one free parameter $a_0$, 
fitted to all galaxies once a particular form for $\mu(x)$ has been chosen.

In the $\Lambda CDM$ cosmological model, the parameter $a_0$ does
not appear.
There are three recent papers pointing out that a shift of paradigm is 
required to describe the data more precisely
\cite {KroupaM}, \cite {FamaeyM}, \cite {McGaughM}.
Kroupa et al. and Famaey et al. discuss simulations based on
the standard cosmological model $\Lambda CDM$. 
They both point out the prediction that the main Dark Matter halo 
hosting a large galaxy like the Milky Way should produce 100--600 roughly 
isotropic sub-halos.
However, the Milky Way has only 24 satellites (and Andromeda $\sim 28$ \cite {Collins}) 
which are highly correlated in both radial and momentum phase space.
They lie mostly within rotationally supported thin discs. 
This has been strongly confirmed by Ibata et al. \cite {Ibata}.
These authors all conclude that $\Lambda CDM$ is at best incomplete and
misses essential physics. 
There is a further interesting paper of L\" ughausen et al. concerning a
polar ring galaxy \cite {Lughausen}.
This galaxy has a small bright gas-poor disc with a large central bulge, but in 
addition an orthogonal gas-rich disc, referred to as a polar ring.
They show that the observed velocities in both discs are well predicted by MOND.

Another $\Lambda CDM$ prediction is that dwarf galaxies formed from tidal 
material during galaxy encounters cannot contain substantial amounts of 
Dark Matter; MOND fits them well.
McGaugh and Milgrom \cite {McGaughM} make an important comment that tidal effects 
of large galaxies on their satellite galaxies must be taken into account in 
drawing conclusions about the satellites.
The objective of the present paper is to search for an explanation of
the observed discrepancies.

The conventional $\Lambda CDM$ paradigm is that Dark Matter condenses 
gravitationally and then galaxies form inside this condensate.
This is a two-step process.
The procedure adopted here is to use commonly used forms of Milgrom's
$\mu$ function to determine the non-Newtonian component of the acceleration
observed at the edges of galaxies; 
it peaks at or close to $a_0$ where it is bigger than $g_N$ by a large factor 
$\sim 171$.
This acceleration is then integrated to determine the associated energy function. 
The result fits naturally to a Fermi function with the same negative sign as that
of gravity.
The Fermi function lowers the total energy by $0.5\, GM$ at radius $r_0$ where 
$g_N$ reaches $a_0$; here $M$ is the mass within radius $r_0$.
This Fermi function takes the same form as the energy gap observed in doped
semiconductors.
It is adopted as the effective Hamiltonian for the non-Newtonian interaction.
It is interpreted as evidence that a Fermi-Dirac condensate forms in the
graviton-nucleon interaction (not the gravitational interaction itself) near 
radius $r_0$.
There are then four further clues which fit like a glove to the existence of 
this condensate.
One is that $2\pi a_0 = cH_0$ within experimental errors; here $c$ is the
velocity of light and $H_0$ is the local value of the Hubble acceleration.
This results in a direct relation between $a_0$ and Dark Energy.
Dark Matter is no longer needed in this one-step process.

Most discussions of galactic rotation curves use the
asymptotic form of the acceleration.
Here attention is focussed on accelerations close to $a_0$, between 
$\sim 10^ {-8}$ and $\sim 10^{-12}$ m\, s$^{-2}$. 

An independent observation of the role of $a_0$ has appeared in globular 
clusters within the last few years.
These spherical clusters of stars have dimensions of a few parsec
(pc), i.e. a factor $\sim 10^4$ smaller than the Milky Way.
They are believed to be the remnants of dwarf galaxies, some old
and some quite young.
Scarpa et al. reported initially on two globular clusters situated 16--19 kpc 
from the Milky Way \cite {Scarpa}.
The equilibrium of such clusters is controlled by Jeans' Law.
Scarpa et al. traced the velocity dispersion of 184 stars at large
radius, identified as being members of one globular cluster (rather than 
interlopers), and 146 stars in the second cluster.
The velocity dispersion is maximal at the centre of each cluster.
They traced it to a radius twice that where $g_N$ reaches $a_0$.
Velocity dispersions deviate rather abruptly from $g_N$ as it decreases 
through $a_0$.
Tidal heating by the Milky Way varies as $r^{-3}$, and is at least one 
order of magnitude smaller, making its effect negligible.

Scarpa et al. have made observations of a further 6 globular clusters.
Hernandez and Jim\' enez give the algebra relating velocity dispersions
of stars to Newtonian acceleration using Jeans' Law 
\cite {HernandezA}.
Hernandez, Jim\' enez and Allen report a detailed study of the velocity
dispersion profiles of all 8 globular clusters \cite {HernandezB}.
Like Scarpa et al., they conclude that tidal effects are significant only
at radii larger by factors 2--10 than the radius where MOND flattens
the curves.
They also show that the velocity dispersion $\sigma$ varies with the mass
$M$ of the cluster as $M^{-4}$ within errors; this is the expected
analogue of the Tully-Fisher relation arising from Jeans' Law.
This result is independent of luminosity measurements used in 
interpreting galactic rotation curves.
In galaxies, the mass $M$ within a particular radius is not easy to
determine, and is usually taken as the mass where rotation curves
flatten out.
Further study of globular clusters is desirable. 

Differences between the $\mu$ functions used for MOND are illustrated in Fig. 19 
of the review of Famaey and McGaugh \cite {Famaey}.
The smoothest form, given by Milgrom \cite {MilgromC}, is used here:
\begin {equation}
\mu (\chi) = \sqrt {1 + 1/(4\chi^2)} - 1/(2\chi).
\end {equation}
The algebraic form with which globular clusters are fitted is closely 
consistent with this equation. 

The layout of this paper is as follows.
Section 2 presents in Fig. 1(a) the observed acceleration on the logarithmic
scale of $g_N$ adopted by Milgrom \cite {MilgromC}; it gives the algebra
for the chosen form of $\mu (x)$, where $x = \log _{10} g_N$.
From this, the difference between the observed acceleration and $g_N$ is
derived and shown in Fig. 1(b).
This is then integrated to find its associated energy function $W(x)$, Fig. 1(c). 
I follow the convention that the zero of the Newtonian potential is taken  
locally, ignoring the effects of Dark Energy over the Universe as a whole. 
The height of the Fermi function may be interpreted as an energy gap.

There is further evidence of quantum mechanics at work.
The asymptotic form of the total acceleration can likewise be integrated.
In Section 3, this is shown to lead to a weak logarithmic tail to the
Newtonian potential. 
The interpretation is that quantum mechanical mixing between
the Newtonian potential and the asymptotic form allows the wavelength of
gravitons trapped in the Newtonian potential to expand. 
This lowers the zero-point energy.

These two indications of quantum effects should not be surprising.
Hawking has demonstrated that Quantum Mechanics plays an important role in 
the physics of Black Holes \cite {Hawking}, where the acceleration is very
large.
A standard Particle Physics result is that the graviton-nucleon interaction
should obey a dispersion relation as a function of the amplitude of $g_N$
(with a non-analytic logarithmic term in addition).
Gravitons with a wavelength of galactic dimensions are at the
extreme infra-red end of the spectrum.
With such a wavelength, a single graviton interacts coherently with
nucleons over complete clusters of stars. 

Section 3 examines the idea that there is quantum mechanical
mixing between $g_N$ and the Hubble acceleration.
This mixing is a well known effect in Particle Spectroscopy for two 
eigenfunctions having different basis states.
The standard treatment is to use the Bogoliubov-Valatin transformation
\cite {Bog} \cite {Valatin}, which leads to the Breit-Rabi 
equation for the mixing. 

Section 4 relates the observed asymptotic acceleration to Dark Energy and a  
subtle connection with the Hubble acceleration.
Subsection 4.1 then draws an analogy with Condensed Matter Physics and 
the well understood Symmetry Breaking in Particle Physics. 
Section 5 makes extensive comments on avenues for further work and
Section 6 summarises conclusion.

\section {\small {THE MODEL NEAR $a_0$}}
From equation (6), 
\begin {eqnarray} 
\mu = g_N/a &=& \sqrt {1 + \frac {a^2_0}{4a^2}} - \frac {a_0}{2a} \\
( g_N/a + a_0/2a) ^2 &=& 1 + (a_0/2a)^2 \\
g_N^2 + a_0g_N &=& a^2.
\end {eqnarray}
The cancellation of the two terms ${a_0/2a}^2$ is typical of a dispersion
relation and emerges as an important point in Section 3.

\begin{figure}[htb]
\begin{center} \vskip -15mm
\epsfig{file=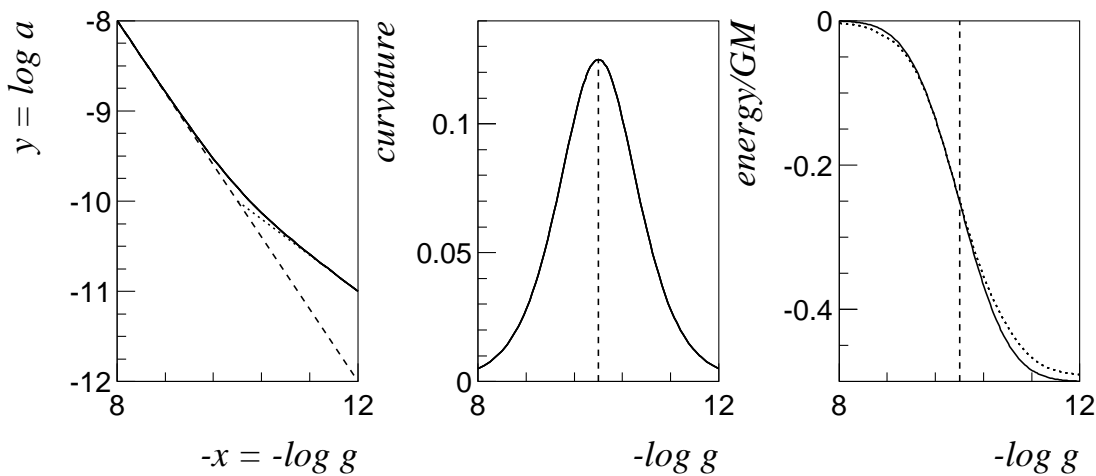,width=16cm}
\vskip -6mm
\caption
{(a) The full curve shows the result of equation (10); the dashed  
line shows $g_N$, and the dotted one a straight line given by $\sqrt {a_0g_N}$;
(b) the peak arising from curvature of the full curve in (a);  
(c) full curve: the energy  derived from (b); the dashed curve is discussed in
Section 3; vertical dashed lines mark $a_0$.}
\end{center}
\end{figure}

Milgrom plots the log of total acceleration against the log of 
gravitational acceleration for systems of different masses \cite {Milgrom9}. 
It is inconvenient to use axes to the base 10, so they will be
replaced in the algebra which follows by conversion to $\rm {\ln}_e$. 
From Eq. (9) 
\begin {eqnarray}
y  & = & \ln _e \sqrt {g_N^2 + a_0g_N}; \\
x  & = & \ln _e \, g_N.
\end {eqnarray}
From equation (10),
\begin {eqnarray}e^{2y} &=& e^{2x} + a_0 e^x \\
dy/dx &=& (e^x + a_0/2)(e^x + a_0)^{-1} \\
d^2y/dx^2 &=& (a_0/2)e^x(e^x + a_0)^{-2} \\
d^3y/dx^3 &=& (a_0/2)e^x(a_0 - e^x)/(e^x + a_0)^{-3};
\end {eqnarray}
$d^3y/dx^3$ goes to zero at $g = a_0$.
[There are higher order terms too.]
The curvature $d^2y/dx^2$ has a maximum value at $x = a_0$, where
$d^2y/dx^2 = 1/8$.

In galaxies, perturbations arise from thermal and pressure 
effects. 
Some narrow structures are observed in rotation curves, for example from 
bars in large galaxies. 
These may be fitted empirically using Poisson's equation for 
mass structures appearing in the gravitational potential.
The MOND component `rides' such structures smoothly, see Figs. 21 and 29
of Famaey and McGaugh \cite {Famaey}.
An immediate question is why Planck's constant does not appear in 
results for galactic rotation curves. 
The reason is that galaxies are noisy enough to hide it.

Using $a_0 = cH_0/2\pi$, the maximum curvature is expected at 
$a_0 = (1.113 \pm 0.046) \times 10^{-10}$ m $s^{-2}$, 
i.e. $\log _{10} a_0 = -9.953$.
McGaugh summarises a large number of papers comparing the Tully-Fisher 
relation with models of galaxy formation \cite {McGaugh11}.
He concludes that gas rich galaxies give the best determination of
the baryonic masses of galaxies: $a_0 = (1.3 \pm 0.3) 
\times 10^{-10}$ m\, $s^{-2}$.
A slightly lower value $1.22 \pm 0.33 \times 10^{-10} \, {\rm m}\, 
s^{-2}$ is found by Gentile, Famaey and de Blok \cite {Gentile}. 
Fig. 1(b) shows $d^2y/dx^2$ as a peak $dW/dx$ in the acceleration at 
$a_0$, marked by the dashed line;
here $W$ is the energy of the `extra' contribution.
Results are insensitive to the precise value of $a_0$, so figures and
arithmetic are simplified by setting $a_0 = 10^{-10}$ m s$^{-2}$.
The full curve is well approximated by a Gaussian for the acceleration:
\begin {equation}
d^2y/dx^2 = 0.125\exp -[\gamma (x - a_0)^2]
\end {equation}
with $\gamma = 1.175$,
i.e. the Gaussian drops to half-height at $9.6\%$ of the value of $x$
at the peak in Fig. 1(b).
The conclusion is that galaxies have considerable stability.

\subsection {Alternative forms for $\mu$}
Other forms for $\mu (\chi)$ have been used to fit galactic rotation 
curves.
Two common examples are (A) $\mu (\chi) = \chi /(1 + \chi)$, 
(B) $\chi /\sqrt {1 + \chi ^2}$.
Both forms have the effect of moving the peak of the curvature slightly: 
from $x = -10$ to $-9.7$ for A and to $-9.85$ for B.
The height of the peak for A is scaled by a factor 0.78 and the curve
becomes correspondingly wider, resulting in a tail reaching 0.012 at
$x = -8.2$ and $-11.8$. 
The result for Fig. 1(c) is that the top of the Fermi function is 
$4\%$ of $GM$ lower, and the bottom of the curve higher by the same amount, 
but the central part of the Fermi function is unchanged and still centred very
close to  $x = -10$.
Form B produces the converse effect: a higher, narrower peak in Fig. 1(b)
and a Fermi function beginning closer to the top and finishing closer to
the bottom of Fig. 1(c), but with its central section undisturbed.

A detail is that one might consider the possibility that the smooth curve
of Fig. 1(a) could be derived from the dashed and dotted lines.
This would give rise to a sharp cusp in Fig. 1(b).
Such cusps are known in Particle Physics, but arise only at the
opening of phase space for a new reaction channel \cite {bugg}.
However, no such threshold exists in galactic phenomena.

\begin{figure}[htb]
\begin{center} \vskip -16mm
\epsfig{file=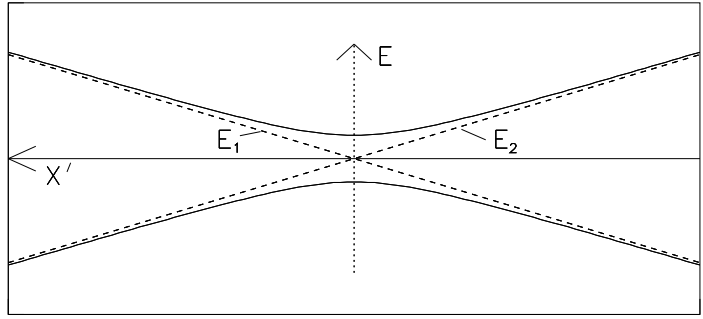,width=13cm}
\vskip -10mm
\caption
{Sketch of two crossing atomic lines; full lines show them including
mixing, dashed lines without; $E_1$ and $E_2$ label the convention for 
eigenvalues in the case of no mixing.}
\end{center}
\end{figure}

In Fig. 1(a), the dashed line corresponding to $g_N$
intersects at $x = -10$ with the straight line fitted to $\sqrt {a_0 g_N}$.
This suggests that there is a cross-over between two eigenfunctions
corresponding to two asymptotic regimes.
This is a familiar result in Particle Physics.
Historically, it was first observed in variations of atomic energy levels
in a magnetic field. 
It arises there from differences in magnetic moments for different levels, some
of which share one component.
The result for that case is sketched in Fig. 2.
The two dashed lines show the variation of energy eigenstates with magnetic
field.

\newpage
For the two mixed states,
\begin {eqnarray}
H_{11}\Psi_1 + V\Psi_2 &=& E\Psi_1 \\
H_{22}\Psi_2 + V\Psi_1 &=& E\Psi_2.
\end{eqnarray}
where $V$ is the mixing Hamiltonian. 
The solution of these coupled equations is
\begin {equation}
(H_{11}-E)(H_{22}-E)-V^2 =0.
\end {equation}
This equation was first derived in 1931 by Breit and Rabi \cite {Breit}.
The same formalism describes mixing between the three neutrinos
$\nu _e$, $\nu_\mu$ and $\nu _\tau$ and also the CKM matrix of QCD.
For galaxies, classical expectation values $<H_{11}>$ and $<H_{22}>$ are to be
substituted into Eq. (19).

\begin{figure}[h]
\begin{center} \vskip -14mm
\epsfig{file=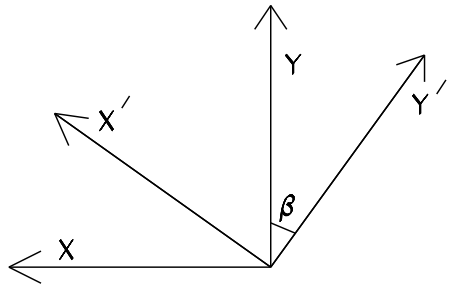,width=11cm}
\vskip -47mm
\caption
{Axes $x$, $y$,\, $x'$ and $y'$.}
\end{center}
\end{figure}
The relevance to galactic physics is that Fig. 1(a) includes mixing between
Newtonian gravity (to the left of the figure) and the Hubble acceleration 
(to the right).
In order to reproduce the symmetric form of Fig. 2, the Bogoliubov 
transformation is needed. 
It rotates Fig. 1(a) by $35.78^\circ$ anti-clockwise; 
this is the mean angle of the dashed and dotted lines with respect to the 
$x$-axis. 
The rotation is illustrated in Fig. 3; it is about the point $x = -10$, 
$y = -10$, where the two straight lines of Fig. 1(a) cross:
\begin {eqnarray}
x' &=& (x+10)\cos \beta - (y + 10) \sin \beta \\
y' &=& (x+10)\sin \beta + (y+10)\cos \beta.
\end {eqnarray}
Substituting Eq. (10) gives an exact expression for the curve in $x',y'$ axes.
 
It is also convenient to re-express $g_N$ and the Newtonian energy directly
in terms of $x$; from Fig. 1(a)
\begin {eqnarray}
H_{11} &=& E_1 = -GM/r = -\sqrt {GM}e^{x/2}\\
H_{22} &=& E_2 = \sqrt {GM}\epsilon (x) \\
V      &=& \sqrt {GM}W(x'),
\end {eqnarray}
where $\epsilon$ refers to an energy possibly given by the Hubble acceleration;
however, in practice, the effect of the Hubble acceleration over the radius
of the Milky Way is $< 2 \times 10^{-4}$, which can be neglected.
Note that the factor $\sqrt {GM}$ appears in all three equations, so as to
conform with Milgrom's scaling law. 
The common approximation is also used that the gravitational energy of a disc
galaxy is $-GM/r$, where $M$ is the mass inside radius $r$.
This is accurate to $\le 1\%$ at the large values of $r$ of interest.
The two solutions of the Breit-Rabi equation are
\begin {equation}
E = \frac {E_1 + E_2}{2} \pm \sqrt{\left( \frac {E_1 - E_2}{2} \right) ^2 +
V^2}.
\end {equation}

\section {\bf The relation to a Fermi function}
The acceleration differs in $x$ and $x'$ axes, but the scalar quantity $W$ is
independent of axes.    
The standard form of the Fermi function is
\begin {equation}  
W(x) \propto \left[ 1 + \exp \left(
\frac {E - E_F}{\beta E_F}\right) \right]^{-1},
\end {equation}
where $E_F$ is the energy at the centre of the Fermi function and $\beta$ is a
fitted constant.
This refers to a situation where there is a discrete energy gap, e.g. in a 
perfect semi-conductor or superconductor.
$W(x)$ can be obtained by numerical integration of $y$ from Eq. (10) and is 
shown as the full curve in Fig. 1(c). 

\subsection {A long-range logarithmic tail to the Newtonian potential}
If there were no quantum mechanical mixing between Newtonian acceleration
and the condensation mechanism, there would be no structure at the crossing
point and, more serious, no explanation for the term $\sqrt {a_0g_N}$. 
 
Asymptotically, the total acceleration, taken from MOND, is
\begin {equation}
a = a_0 \sqrt {g_N/a_0}.
\end {equation}
Since $g_N = GM/r^2$,
\begin {equation}
a \to \sqrt {GMa_0}/r.
\end {equation}
Taking this as $-d\phi /dr$, where $\phi$ is a potential induced
by the mixing,
\begin {equation}
\phi \to -\sqrt {GMa_0}\ln (r/r_1).
\end {equation}
Here $r_1$ is the mean radius for this
term. 
The value of $r_1$ is necessarily very close to the much larger
dip caused by $dW/dx$.
Because $a_0 \sim 10^{-10}$, $\phi$ is very small.
However, it does explain the asymptotic straight-line at the right-hand
edge of Fig. 1(a) and contributes to $\epsilon (x)$ in $H_{22}$.

The interpretation of this term is that mixing between the Newtonian potential 
and the condensation mechanism expands the wave-length of gravitons trapped 
in the Newtonian potential and lowers their zero-point energy.
An analogy is the covalent bond in chemistry.
In the hydrogen molecule, each of two electrons is attracted to two protons 
(which themselves repel one another).
This increases the wave-length of the electrons and reduces their zero-point 
energy. 

Sobouti noticed this long-range logarithmic
component and wrote two closely related papers \cite {SiboutiA},
\cite {SiboutiB}. 
These included effects of General Relativity.
This is interesting but not strictly necessary at present.
The problem was solved by Sobouti as a power series with additional
empirical terms proportional to $1/r$ and $1/r^2$.
These terms are now replaced by our equations.
There is experimental evidence that the Milky Way has a halo extending to 
$\sim 100$ kpc \cite {Deason}.
However, this could be due to gas and dust shared with the local cluster
of galaxies.
So this halo is presently ambiguous.

\subsection {Solving the Breit-Rabi equation}
The binding energy $W$ is defined with the same sign as 
$E_1$. 
From equations (25) and (24), 
\begin {eqnarray}
2E &=& E_1 + \epsilon (x) \pm 
\sqrt {(E_1-\epsilon (x))^2  + 4V^2} \\
&=& \frac 
{\sqrt {GM}}{\rm {ln} _e {10}}
\left( -e^{x/2}+\epsilon (x) 
\pm \sqrt {(e^{x/2}-\epsilon (x))^2 + 4W^2}\right) \\
2\frac {dE}{dx} &=& \frac 
{\sqrt {GM}}{\rm {ln} _e {10}}
\left( -0.5e^{x/2} +d\epsilon /dx \pm 
\frac {(e^{x/2}-\epsilon (x) )(0.5e^{x/2}-d\epsilon (x)/dx) + 4 WdW/dx}
{\sqrt {(e^{x/2} - \epsilon (x))^2 + 4W^2}} \right)
\end {eqnarray}
The factor $\rm {ln} _e {10}$ allows for the fact that
Fig. 1(a) has been drawn using axes which use logarithms to the base 10.
For the upper branch of the solution, the minus sign for the term involving
the square root is required to reproduce the usual Newtonian potential.
The equations simplify considerably if the very small term $\epsilon (x)$ is
neglected. 

A Bose-Einstein condensate does not fit the data. 
In this case, $dW/dx$ should vary as $T^{3/2}$ at the peak of the 
acceleration \cite {Tilley2}.  
For positive $x$ near $x = 0$, the relation of the energy function to $kT$ gives,  
\begin {eqnarray}
W(x) &=& -B(1 + |x|\,^{3/2}\exp {-\gamma '\, x{^2}}) \\
dW(x)/dx' &=& B(1.5|x|\,^{0.5} - 2\gamma '\,|x|\,^{5/2})
\exp {-\gamma '\, x\,^2}; 
\end {eqnarray}
for negative $x$, the opposite sign of $|x|$ is needed in $W(x)$.
The second term in $dW(x)/dx$ wrecks the $x$ dependence, which fails
to fit the observed peak.
Near $x=0$ it has a square root variation with $x$ and then, when
the second term of Eq. (34) overtakes it, the curve turns downwards. 
This rules out a Bose-Einstein condensate.

The depth of the Fermi function is $-0.5\, GM$.
The magnitude of this term can be traced to the factor 2 difference
in slope of $g_N$ and that of the asymptotic form $\sqrt {a_0 g_N}$.

Let us now return to Figs. 1(b) and (c).
Here there is a slight complication. 
Fig. 1(b) is the acceleration measured in $x,y$ axes. 
However, the rotation to $x',y'$ axes requires that $\gamma$ of Eq. (16)
is increased to 1.852.
In addition, there is a small visible displacement of the centre of curvature
in Fig. 1(a) by an offset of $-0.203 \times 10^{-10}$ in $x$. 
What then emerges from the Breit-Rabi equation is that 
$g_N$ is rather small near $x = a_0$ compared with that originating from 
the extra acceleration $dW/dx'$.
This second term dominates by a large factor $\sim 171$ at the 
centre of the curve.
This ratio falls by $50\%$ at $x=-10.6$, to 30 at $x=-11$, then -6.0 at 
$x=-11.5$ and  $\sim 1$ at $x=-12$.  
Results for the `extra' acceleration are symmetric about $a_0$ except for 
the term $\sqrt {a_0g_N}$ of Eq. (10).
The conclusion is that the curved part of Fig. 1(a) is the dominant 
feature near $x = a_0$ where Newtonian gravitation is a rather small 
perturbation.

Consider the effect of this result near the centre of the Fermi function 
at $E = E_F$ in Fig. 1(c). 
If we retain only the dominant terms in $W$ and $dW/dx$, Eqs. (31) and 
(32) give
\begin {eqnarray}
dE/dx &\to& \frac {\sqrt {GM}}{ln_e {10}}2dW/dx \\
E     &\to& \frac {\sqrt {GM}}{ln_e {10}}2W.
\end {eqnarray}
Apart from the factor $\sqrt {GM}/ln_e {10}$, which is a normalisation factor,
$dE/dx$ may be interpreted as the modulus of a Breit-Wigner resonance with
$x$-dependent width:
\begin {equation}
BW = \frac {\Gamma (x)/2}{E - E_F - i\Gamma (x)/2}.
\end {equation}   
The energy $W$ starts at zero because of local gauge invariance, and its
central value is shifted downward by 0.25 GM.
This accounts for the form of Eq. (8) of Section 2, where two terms
$(a_0/2a)^2$ cancel.
 
How can this effect be understood?
A possibility is that the graviton acquires a small effective mass 
near $a_0$.
The fitted change to the gravitational acceleration is close to
a Gaussian, as in Fig. 1(b).
In subsection 2.1, the dependence of the fit on alternative forms of
Milgrom's $\mu$ function was tested.
Although acceleration curves change significantly, the Fermi function
is affected only at the ends of the range $x=8$ to $12$ by at most
$\pm 4\%$.

The conclusion from these results is that the central part of the Fermi
function is stable, but can be perturbed at the edges.
In superconductors, a coherence length was introduced by Pippard to
account for the effects of defects beyond an experimentally observed range
\cite {Pippard}.
It appears that galaxies behave similarly.
If gravitons acquire an effective mass, it appears at first sight that 
this will weaken gravity.
However, remember that the wave-lengths of gravitons on a galactic
scale are very large.
The spectrum of gravitons near the edge of the galaxy is a convolution
of gravitons from the rest of the galaxy. 
Gravitons from the galactic centre become almost plane waves 
which can interact coherently over a large volume.
They function like phonons coupling to Cooper pairs in a superconductor 
producing an energy gap.
A coherence length like that introduced by Pippard can arise in many ways.
Supernovae act as major perturbations, heating sizable volumes.
It is also known that so-called chimneys and wormholes provide
channels through which currents of dust and gas flow.
Furthermore, one of the remarkable features of galaxies is that they only 
grow to a certain size.
The largest have masses of order $10^{12} M_{\odot}$.
This can be attributed to the rapid fall-off of the `extra' acceleration
due to MOND close to $x = -8$ and $-12$.
All of these effects point to a coherence length due to
the variation of the Fermi function in Fig. 1(c).

Let us return to a simpler issue, the missing lower branch of the
Breit-Rabi equation. 
On this branch both $W$ and $dW/dx'$ change sign.
The change of sign requires that this branch describes an excited
state rather than a condensate. 
(Remember that energies of both gravity and $W(x)$ are negative). 
Such an excited state is likely to decay on a time scale much less than
that of galaxies, so it is unlikely that this branch will be observable.
For those wishing to investigate this branch, 
the procedure is (a) to fit the upper branch as a function of $x$,
(b) rotate to $x',y'$ axes, (c) reverse the sign of $y'$ to reach the
lower branch, and (d) rotate axes back again to $x,y$.
The best place to search for this branch is near the crossing point
of Fig. 1(a).

\section {The relation to Dark Energy}
Experiment tells us that in galaxies, the asymptotic form of the
acceleration is $\sqrt {a_0g_N}$.
This leads to the question: what governs the asymptotic acceleration? 

If MOND successfully models the formation of galaxies and globular
clusters, it raises the question of how to interpret Dark Energy.
In a de Sitter universe, the Friedmann-Robertson-Walker model  
smoothes out structures using a $\Lambda CDM$ function which models the 
gross features.
These change over the lifetime of the Universe.
However, if quantum mechanics governs individual galaxies, there will
instead be fine structure in Dark Energy. 
It is logical that steps like Figs. 1(c) do not just average out, but 
instead accumulate over all of this fine structure.

The standard treatment of Dark Energy is reviewed by Sami \cite {Sami}.
The acceleration falls initially due to Newtonian cosmology. 
However, at recent times, the acceleration increases. 
This acceleration can be explained naturally by the sum total of the fine 
structure over all the galaxies.
The total acceleration is parametrised via the assumed time dependence of the 
metric on the Hubble acceleration.
A similar suggestion along these lines has been advanced by Zhang and Li
\cite {Zhang} using ideas based on entropic arguments. 
The `present-day' Hubble acceleration is the local value and varies over
the Universe according to the parametrisation by Dark Energy.

There is a further argument pointing towards the idea that local fine
structure is cumulative.
Peebles and Nusser argue that galaxies condense more rapidly than the 
standard $\Lambda CDM$ model predicts \cite {Nusser}.
In particular, they point out that the Local Void contains far
fewer galaxies than $\Lambda CDM$ predicts statistically,
while there is an unexpected presence of large galaxies on the
outskirts of the Local Void.
Their Fig. 1 is very persuasive in this respect.
Only 3 galaxies are observed in the Local Void compared with 19
predicted.  
The Poisson probability for this result 
is of the order $10^{-5}$ from $\Lambda CDM$.
Peebles and Nusser conclude: `In short, the general sensitivity of
galaxies to their environment is not expected in standard ideas.
It would help if galaxies were more rapidly assembled so that they
could then evolve as more nearly isolated island universes.'
Later Peebles considered an additional empirical term added to the
$\Lambda CDM$ model, but commented that the change requires that
Cold Dark Matter is cored rather than the expected cusped behaviour
\cite {Peebles}.

A natural explanation is that the Local Void gives no contribution
to the Hubble mechanism, except for its three galaxies.
The total energy $E$ is then higher there.
On the periphery of the Void, there is gas and dust which can form
galaxies.
This gas runs down the energy gap to enlarge galaxies forming there.
Galaxies then grow by accumulation of dust and gas.
The condensate at the edge of the galaxy acts as a funnel to collect
gas and dust.
Towards the centres of galaxies, they can develop in different ways
depending on the angular momentum $L$ of the galaxy.
For large $L$, they naturally form flat discs.
For lower $L$, bulges develop and spiral arms; these features can vary 
according to whether the galaxy is supported by rotation or pressure or 
both.
For really low $L$, large elliptical galaxies and dwarf spherical galaxies 
develop.
If $L$ is close to 0, galaxies collapse as quasars.
It appears that $\Lambda CDM$ is presently fitting all these different
morphologies with a very flexible ansatz for Dark Matter.

So far, Dark Matter has escaped experimental detection.
There are speculations that it may take the form of sterile neutrinos
with masses in the electron-volt range.
At a recent conference on neutrino physics, Altarelli commented on
this question \cite {Altarelli}.
He argued that more than one sterile neutrino is disfavoured by stringent
bounds arising from nucleo-synthesis.
Also there is tension between LSND, MiniBoone and KARMEN experiments leaving
little room for a signal.
Rubbia et al. propose a new neutrino detector ICARUS-NESSIE to run at CERN with 
parameters optimised for finding sterile neutrinos \cite {Rubbia}.   

There is also speculation about Grand Unified Theories in which massive
right-handed neutrinos couple weakly to left-handed light neutrinos.
Such heavy neutrinos would have formed in the Big Bang before the
generation of the Cosmic Microwave Background.
If these heavy neutrinos survive and mix with light neutrinos, their lifetimes
must be larger than the age of the Universe, otherwise there would be
effects visible in Dark Energy.

\subsection {Other Condensates}
Well known examples of condensates are ferromagnets and anti-ferromagnets.
In ferromagnets, spins align parallel with one another below the Curie point; 
in the absence of a magnetic field, the overall spin can lie in any direction.
In anti-ferromagnets they spontaneously align anti-parallel.
The ground-states of these systems have a lower symmetry than the Hamiltonian.
This is a purely quantum effect, but difficult to calculate from first
principles. 
However, neutron scattering experiments establish the structures.
In astrophysics, the situation is difficult because one can only watch
how galaxies evolve. 
 
In Particle Physics. we now know that the Strong Interaction is mediated by 
massless gluons obeying Chiral Symmetry, i.e. they do not discriminate 
between left-hands and right-hands. 
Below a mass of $\sim 1 $ GeV, the gluon acquires an effective mass from its 
interaction with light quarks, which themselves have masses of $\sim 4$ and 9 
MeV.
The Electroweak Theory is constructed from a mixing between Electromagnetism
and Weak Interactions carried by $W$ and $Z$ particles and probably the
Higgs boson. 
A side-effect of this idea is that Chiral Symmetry is broken in the
strong interactions for spin 0 particles below 1 GeV. 
This idea was introduced by Gasser and Leutwyler \cite {Gasser} and today gives 
many quantitatively accurate results in meson spectroscopy. 
An important step in understanding the precise mechanism was made by 
Bicudo and Ribiero \cite {Bicudo}.
Their work finds the Bogoliubov-Valatin transformation to be an
essential element.
Above 1 GeV, there is a cross-over in which Chiral Symmetry is largely
restored and the quark model reigns supreme, though with small amounts
of mixing with meson-meson and/or $q\bar q q\bar q$ basis states.
A recent paper of Pennington and Wilson gives details including figures 
showing the cross-over at 1 GeV \cite {Pennington}.

A precise set of equations describing Chiral Symmetry Breaking is given by
Cherney et al. \cite {Cherney}.
This work shows that gauge invariance requires that quarks 
must be treated as `dressed' fermions, rather than bare fermions.
Their conclusion is that a Yukawa potential appears explicitly in 
the $q\bar q$ interaction,  their Eq. (61). 
This increases the fermion mass from a few MeV to the so-called constituent 
mass, $\sim 320$ MeV.
In their Eq. (72), a logarithmic term appears from Feynman diagrams
where one pair of particles rescatter to themselves through a closed
loop.
The origin of Chiral Symmetry Breaking is that quarks have small masses;
if these were zero, the pion would be massless. 

In summary, it is clear that there are three gauge fields: (i) gluons, (ii) 
electromagnetism linked to weak interactions, and (iii) gravity.
Effective masses do arise in the first two.
There has been a paper by Van Dam and Veltman \cite {Dam} claiming a no-go 
theorem preventing the graviton developing a mass. 
This theorem appears to refer purely to an isolated graviton obeying General
Relativity.
That is different to the present case where the condensate is a property of
the graviton-nucleon interaction.

There is an important point if gravitons develop an effective mass.
In Particle Physics, there are relations between scattering processes  such as
$\pi ^+p \to \pi^+p$, $\pi ^-p \to \pi^-p$ and $\pi ^+\pi ^- \to p\bar p$.
Amplitudes depend on $E^2 - p^2$, where $E$ and $p$ are momenta in each
reaction. 
These are denoted $s$, $u$ and $t$-channels by a convention introduced 
by Mandelstam. 
For massless gravitons, kinematics of the $s$-, $t$- and $u$- channels all 
meet at a point.
However, if the graviton acquires an effective mass, it moves these three
channels apart and changes the graviton-nucleon interaction.
In particular, it affects the nucleon pole term in the $u$-channel,
hence the structure of the nucleon.
What is needed, but beyond reach at present, is to solve the
Schwinger-Dyson equations for this case following the procedure
of Pennington and Wilson for the $\pi \pi$ system.
The obstacle is that the structure of the nucleon is not yet known
precisely enough for a realistic calculation from first principles.
Thomas developed a model in 1983 explaining features of deep inelastic
scattering (i.e. large momentum and energy transfers in electron
scattering from nucleons) which provided a qualitative explanation
of the data in terms of a chiral quark model \cite {Thomas}.
The idea is that a nucleon has a component in its wave function with a pion 
circulating round it with one unit of orbital angular momentum: 
$N \to [N+\pi]_{L=1}$.
This has been followed up in a recent paper of Burkhardt et al. which
improves this work in specific ways \cite {Burkhardt}, but the final set
of parameters is not yet available.

Cosmology needs to be treated equally empirically.
My suggestion is that Dark Energy is symptomatic of spontaneous symmetry breaking
of gravitation to a de Sitter universe governed by the space and time components
of the metric.
The de Sitter universe has a lower symmetry than General Relativity.
In galaxies, $a_0$ is the order parameter of a Fermi-Dirac condensate and is 
directly related to $cH_0/2\pi$.
     
\section {Further work which is needed}
The model proposed here is precise and open to experimental test. 
The most important and simplest is that if Dark Matter is replaced by 
this model, it is obviously necessary to redo the parametrisation of Dark 
Energy so that it reproduces smoothly what has been parametrised as Dark 
Matter up to now.
This does not necessarily require major modifications.
The main point is to fit the third peak in the Cosmic Microwave Background.
This requires cooperation of groups with the latest Planck data and techniques 
at their finger-tips.
Despite the clues pointed out here that a Fermi-Dirac condensate explains 
galactic rotation curves, this could fail.
It would not be surprising if minor modifications are required
in the refit to Dark Energy and new clues might emerge.
It is necessary to include the logarithmic tail of the Newtonian potential
found here. 
It is also essential to treat Voids so that they only contribute to
Dark Energy via the galaxies observed there.
Those may not be well resolved if they are very distant from us.
It is also essential to fit the Hubble parameter and magnitudes of Type1a
supernovae as a function of red-shift $z$.

Whatever develops at the edges of galaxies  will affect both Strong and Weak 
Gravitational Lensing. 
However, unless data are used where a distant quasar transmits light 
through the periphery of a galaxy, where the effect of $W(x)$ is large, 
the lensing effect will be rather small.
The weak logarithmic tail of Newtonian gravitation also needs to be taken 
into account.

What happens in clusters of galaxies needs detailed, laborious
calculations using the formulae given here; since there are presently
claims that Milgrom's formula does not correctly reproduce what happens
in clusters, this could be revealing.
A multi-body interaction between the Hubble mechanism and several 
galaxies is required.
Quite apart from the attraction to acceleration $a_0$, 
there are also strong tidal effects of the variety discussed by Kroupa 
\cite {Kroupa2}.
 
The same remarks apply to the Bullet Cluster, which needs to be refitted 
with the model proposed here.
The calculation becomes a two-centre problem. 
The Hubble acceleration couples to both galaxies, but they also couple to 
one another, modifying the zero-point energy;  
individual stars in galaxies communicate with one another not only
through the Newtonian potential but via their Fermi functions.
In the Bullet Cluster, each galaxy behaves in this way, but there will
be coupling between stars "belonging" to each  individual galaxy.
There may be complex interactions between the two galaxies including 
resonance effects. 
There are new data on the Bullet Cluster \cite {Paraficz} showing that
there are many dwarf satellite galaxies in the cluster. 

\section {Concluding Remarks}
The equations given here allow very little freedom - just the $4\%$ 
perturbations allowed in the Fermi function at the top and bottom of
Fig. 1(c). 
There are five clues which point to galaxies and globular clusters being
quantum mechanical condensates.

\begin {enumerate} 
\item The phenomena of Fig. 1 appear on a logarithmic scale of $g_N$.
This is the form expected for the Partition function of Statistical
Mechanics:
\begin {equation}
Z = \Pi_s \frac {1}{1 + e^{E_s/kT}}
\end {equation}
where $E_s$ are energies of levels in a box and $\Pi$ denotes the numbers
of quantum levels in the box.
For a refresher course, see Schr\" odinger's clear exposition \cite {Schrodinger}.
\item 
Fig. 1(a) can be interpreted in terms of quantum mechanical mixing between two 
crossing eigenstates; the rotation of axes accomodated by Fig. 3 is just what is 
expected from the Bogoliubov-Valatin transformation, a quantum effect.
\item Fig. 1(c) is fitted naturally by a Fermi function with 
an energy gap $0.5\, GM$.

\item The asymptotic form of the acceleration in Fig. 1(a) generates
a logarithmic tail; this requires a quantum mechanical explanation.
A Fermi-Dirac condensate fits the data; a Bose-Einstein condensate 
does not.
\item Long wave-length gravitons can explain the amplification of
the amplitude forming the condensate.
\end {enumerate} 

The lower branch of the Breit-Rabi equation has the opposite curvature to 
the upper one. 
The natural interpretation of this result is that it corresponds to an 
excited state which will decay rapidly and will not therefore appear in 
galactic rotation curves.

The interpretation of the curve of Fig. 1(b) as an energy-dependent 
Breit-Wigner pole is that the graviton acquires an effective 
mass in the vicinity of $a_0$.
How this originates is speculative but does not reflect on the five clues 
listed above; it depends on the nucleon structure function which is not yet
well defined.

The asymptotic form of the acceleration is $\sqrt {a_0g}$.
This is clearly associated with the Hubble acceleration and is related
to Dark Energy. 
The standard approach to the Universe is the Friedmann-Robertson-Walker 
model where components of the metric are smoothed over local structures 
and appear in the radial and time components of the metric of General 
Relativity.
My suggestion is that galaxies create fine-structure and the 
Friedmann-Robertson-Walker model should include into Dark Energy the sum 
over these structures.
This may explain the agreement between $a_0$ and $cH_0/2\pi$ in our
part of the Universe as well as its late-time acceleration.

Those are the essential points.  
They do reflect successfully the facts which MOND parametrises. 
However, only the peripheries of galaxies are considered. 
The morphology at the centres of galaxies is a separate issue.  

There are two further small comments.
Firstly, there have been many papers on the derivation of Newton's law from
Entropic arguments. 
Many assume that one can assign a temperature $T_0$ to the Hubble
acceleration and a temperature $T_1$ for the gravitational potential 
by adding them as the sum of squares, i.e. the random phase 
approximation. 
This gives the result for the total acceleration:
\begin {equation}
a = \sqrt {(g_N + g_{H})^2 - g_H^2}
\end {equation}
where $g_H$ is the Hubble acceleration $H_0^2 r$, and $r$ is the
distance to the centre of the gravitating object. 
This is a quite different approach to the one proposed here and
leads to a very small effect. 
For the Milky Way, it is a maximum of $2 \times 10^{-4} cH_0/2\pi$.

Secondly, Iorio has calculated that effects of MOND on the perihelia of 
planets in the Solar system are about a factor 10 below present 
experimental errors \cite {Iorio}. 
The logarithmic term arising from MOND will be of order $a_0$, which
is very small and will vary exceedingly slowly over the solar system.
It has been pointed out by Galianni et al. that it may be feasible to
detect the MOND effect at acceleration $a_0$ in the solar system near 
Lagrangian points where the accelerations of the Sun, Earth and Moon 
cancel out \cite {Galianni}.
If measurements of sufficient accuracy could be made,
they might confirm or modify MOND as a model of the behaviour of galaxies; 
secondly, they have the important potential to measure the shape of the 
response function through the region where the bend appears in Fig. 1. 
Further study of globular clusters would also provide information on the 
rotation curves of galaxies.
\vskip 3mm
\noindent {\small ACKNOWLEGEMENT}
\newline
I wish to thank Prof. Pedro Bicudo for discussions about Bose-Einstein 
condensates.


\begin {thebibliography} {99}
\bibitem {Famaey}      
B. Famaey  and S.S. McGough, arXiv: 1112.3960.
\bibitem {MilgromA}    
M. Milgrom, Astrophys. J {\bf 270} 371 (1983). 
\bibitem {MilgromB}    
M. Milgrom, Astrophys. J {\bf 270} 384 (1983).
\bibitem {Milgroms}    
M. Milgrom, ApJ {\bf 698} 1630 (2009). 
\bibitem {Tully}       
N.B. Tully and J.R. Fisher, Astron. Astrophys. {\bf 54} 661 (1977). 
\bibitem {McGaughA}    
S.S. McGaugh, Astrophys. J {\bf 632} 859 (2005).
\bibitem {KroupaM}     
P. Kroupa, M. Pawlowski and M. Milgrom, arXiv: 1301.3907.
\bibitem {FamaeyM}     
B. Famaey  and S.S. McGough, arXiv: 1301.0623.
\bibitem {McGaughM}    
S. McGaugh and M. Milgrom, Astrophys. J {\bf 766} 22 (2013).
\bibitem {Collins}     
M.L.M. Collins {\it et al.}, arXiv: 1302.6590.
\bibitem {Ibata}       
R.A. Ibata {\it et al.} arXiv: 1301.0446.
\bibitem {Lughausen}   
F. L\" ughausen {\it et al.} arXiv: 1304.4931.
\bibitem {Scarpa}      
R. Scarpa {\it et al.}, Astron. Astrophys. {\bf 525} A148 (2011).
\bibitem {HernandezA}  
X. Hernandez and M.A. Jim\' enez, Astrophys. J {\bf 750} 9 (2012).
\bibitem {HernandezB}  
X. Hernandez, M.A. Jim\' enez and C. Allen, arXiv: 1206.5024.
\bibitem {MilgromC}    
M. Milgrom, Proc. 2nd Int. Workshop on Mark Matter (DARK98) eds. 
H.V.Klapdor-Kleingrothaus, L. Baudis: arXiv: astro-ph/9810302.
\bibitem {Hawking}     
S.W. Hawking, Comm. Math. Phys. {\bf 43} 199 (1975). 
\bibitem {Bog}         
N.N. Bogoliubov,  J. Exptl. Theor. Phys. (U.S.S.R) {\bf 34} 58,73 (1958);

translation:  Soviet Phys. JETP {\bf 34} 41, 51
.
\bibitem {Valatin}     
J.G. Valatin, Nu. Cim. {\bf 7} 843 (1958).
\bibitem {Milgrom9}    
M. Milgrom, arXiv: 0908.3842.
\bibitem {McGaugh11}   
S.S. McGaugh, arXiv: 1107.2934.
\bibitem {Gentile}     
G. Gentile, B. Famaey  and W.J.G. de Blok, Astron. Astrophys. {\bf 527} 
A76 (2011). 
\bibitem {bugg}        
D.V. Bugg, J. Phys. G: Nucl. Part. Phys. {\bf 35} 075005 (2008).
\bibitem {Breit}       
G. Breit and I.I. Rabi, Phys. Rev. {\bf 38} 2082 (1931).
\bibitem {SiboutiA}    
Y. Sobouti, arXiv: 0812.4127. 
\bibitem {SiboutiB}    
Y. Sobouti, arXiv: 0810.2198.
\bibitem {Deason}      
A.J. Deason {\it et al.} arXiv: 1205.6203.
\bibitem {Tilley2}     
D.R. Tilley and J. Tilley,  {\it Superfluidity and Superconductivity},
Adam Hilger, Bristol  and New York, 3$^{rd}$ Edition (1990).
\bibitem {Pippard}     
A.B. Pippard, Proc. R. Soc. A {\bf 216} 547 (1953). 
\bibitem {Sami}        
M. Sami, Curr. Sci. {\bf 97} 887 (2009).
\bibitem {Zhang}       
H. Zhang and X-Z Li, Phys. Lett. B {\bf 715} (2012) 15.
\bibitem {Nusser}      
P.J.E. Peebles and A. Nusser, Nature {\bf 465} 565 (2010).
\bibitem {Peebles}     
P.J.E. Peebles, arXiv: 1204.0485.
\bibitem {Altarelli}   
G. Altarelli, arXiv: 1304.5047.
\bibitem {Rubbia}      
C. Rubbia, A. Guglielmi, F. Pietropaolo, and P. Sala, arXiv: 1304.2047.
\bibitem {Gasser}      
J. Gasser and H. Leutwyler, Phys. Lett. B {\bf 125} 325 (1983);
Annals Phys. {\bf 158} 142 (1984).
\bibitem {Bicudo}      
P. D. de A. Bicudo and J.E.F.T. Ribiero, Phys. Rev. D {\bf 42} 1611 
(1990).
\bibitem {Pennington}  
M.R. Pennington and D.J. Wilson, Phys. Rev. D {\bf 84} 094028 (2011) and
119901 (E) (2011).
\bibitem {Cherney}     
A.Yu. Cherny, A.E. Dorokhov, Nquyen Suan Han, V.N. Pervushin and V.I. Shilin,
arXiv: 1112.5856.
\bibitem {Dam}         
H. van Dam and M.J.G Veltman, Nucl. Phys. B {\bf 22} 397 (1970).
\bibitem {Thomas}      
A.W. Thomas, Phys. Lett. B {\bf 126} 97 (1983).
\bibitem {Burkhardt}   
M. Burkardt, K.S. Hendricks, Chueng-Ryong Ji, W. Melnitchouk and A.W. Thomas, 
arXiv: 1211.5853.
\bibitem {Kroupa2}     
P.Kroupa, arXiv: 1204.2546. 
\bibitem {Paraficz}    
D. Paraficz {\it et al.} arXiv: 1209.0384.
\bibitem {Schrodinger} 
E. Schr\" odinger, {\it Statistical Thermodynamics}, Cambridge University 
Press, 2$^{nd}$ Edition (1952).
\bibitem {Iorio}       
L. Iorio, arXiv: 0905.4704.
\bibitem {Galianni}    
P. Galianni, M. Feix, H.S. Zhao and K. Horne,  arXiv: 1111.6681.
\end {thebibliography}
\end {document}